\documentclass[a4paper,fleqn,final]{cas-dc}

\usepackage[T1]{fontenc}
\usepackage[utf8]{inputenc}
\usepackage[main=english,romanian]{babel}
\usepackage{textcomp}
\usepackage{hyperref}
\usepackage{amsmath}
\usepackage{amssymb}
\usepackage{graphicx}
\usepackage{booktabs}
\usepackage{tabularx}
\usepackage{array}
\usepackage{threeparttable}
\usepackage[numbers,sort&compress]{natbib}
\usepackage{longtable}
\usepackage{adjustbox}
\usepackage{caption}
\usepackage{etoolbox}
\usepackage{placeins}
\usepackage{xurl}
\usepackage{microtype}

\makeatletter
\g@addto@macro{\UrlBreaks}{\do\_\do\-\do\/\do\.\do\0\do\1\do\2\do\3\do\4\do\5\do\6\do\7\do\8\do\9}
\makeatother

\renewcommand{\ttfamily}{\rmfamily}
\AtBeginDocument{%
  \urlstyle{same}%
}

\usepackage{xcolor}
\definecolor{linknavy}{RGB}{0,70,127}
\hypersetup{
  colorlinks=true,
  linkcolor=black,      
  citecolor=linknavy,   
  urlcolor=linknavy,    
  runcolor=linknavy
}

\AtBeginEnvironment{table}{\let\sffamily\rmfamily}
\AtBeginEnvironment{figure}{\let\sffamily\rmfamily}
\AtBeginEnvironment{table*}{\let\sffamily\rmfamily}
\AtBeginEnvironment{figure*}{\let\sffamily\rmfamily}

\ExplSyntaxOn
\RenewDocumentCommand \firstname {}
	{ \textcolor{black}{\seq_use:Nn \l_stm_au_seq { ~ }} }

\RenewDocumentCommand \emailauthor { m m }
   {
     \int_gincr:N \g_ead_int
     \seq_gput_right:Nn \g_stm_ead_seq
       {
         { \href{mailto:#1}{\rmfamily #1} }
         \parsename { #2 }
         \space(\eadauthor)
       }
     }

\cs_set:Npn \__first_footerline:
{
	\group_begin:
	\small
	\normalfont
	\ifnum\theblind>0\relax
	\else
	\__short_authors: :~
	\fi
	\itshape Preprint~submitted~to~Taylor~\&~Francis
	\group_end:
}

\RenewDocumentCommand \printorcid { } { }

\cs_set:Npn \__first_head:
{
	\parbox[t]{\textwidth}
	{
		\rule{\textwidth}{0pt}
	}
}

\cs_set:Npn \__cas_head:
{
	\parbox{\textwidth}
	{
		\rule{\textwidth}{0pt}
	}
}

\cs_set:Npn \__cas_foot:
{
	\parbox[t]{\textwidth}
	{
	 \rule{\textwidth}{.2pt}\\
	 \small
	 \normalfont
	 \__first_footerline:
	 \hfill Page~\thepage {}~of~ \lastpage
	}
}
\ExplSyntaxOff

\makeatletter
\ps@cas
\makeatother

\begin{document}

\makeatletter
\def\bstctlcite{\@ifnextchar[{\@bstctlcite}{\@bstctlcite[@auxout]}}
\def\@bstctlcite[#1]#2{\@bsphack
  \@for\@citeb:=#2\do{%
    \edef\@citeb{\expandafter\@firstofone\@citeb}%
    \if@filesw\immediate\write\csname #1\endcsname{\string\citation{\@citeb}}\fi}%
  \@esphack}
\makeatother
\bstctlcite{IEEEbsTcontrol}

\shortauthors{Narimani et al.}
\shorttitle{Optical GeoAI and urban-canopy screening}

\title[mode=title]{Optical GeoAI and spatial modeling enable urban-canopy mapping and resilience screening in Davis, California}

\author[1]{Mohammadreza Narimani}
\cormark[1]
\ead{mnarimani@ucdavis.edu}

\author[2]{Shreyan Mitra}

\author[1]{Parastoo Farajpoor}

\affiliation[1]{organization={Department of Biological and Agricultural Engineering, University of California, Davis},city={Davis},state={CA},postcode={95616},country={USA}}

\affiliation[2]{organization={California High School},city={San Ramon},state={CA},postcode={94583},country={USA}}

\cortext[1]{Corresponding author}

\begin{abstract}
Timely urban-canopy information is essential for linking remote sensing with heat, mobility, and neighborhood planning. We developed an optical GeoAI workflow for Davis, California, using 2022 National Agriculture Imagery Program imagery (0.6\,m RGB+NIR). DeepForest generated crown candidates; an NDVI threshold, non-maximum suppression, and box-prompted Segment Anything Model (ViT-B) produced a crown-anchored canopy surface. Analyses used the 25.92\,km$^2$ Census TIGER municipal boundary and a 100\,m grid. The workflow retained 11{,}741 candidate crowns and mapped 7.71\,km$^2$ of canopy (29.8\% of the city). On the identical extent, pixel precision was 0.804, pixel recall was 0.873, and 97.4\% of candidate centers agreed with the 2022 USDA/CAL FIRE LiDAR-assisted canopy product (IoU 0.719; Dice 0.837; area recovery 108.5\%). Approximately 49\% of candidates occurred within 15\,m of a road. Canopy was inversely associated with Landsat land-surface temperature (Spearman $\rho = -0.477$; partial $\rho = -0.551$ controlling for built probability), and spatial-lag modeling confirmed clear neighborhood structure. Two transparent attention surfaces combined canopy need with thermal and contextual indicators. The framework provides a reproducible, updateable screening layer that complements structural canopy products and municipal inventories while retaining assumptions, data provenance, and spatial diagnostics for planning interpretation.
\end{abstract}


\maketitle


\section{Introduction}\label{sec:introduction}

Urban trees function as distributed environmental infrastructure. Their canopies alter radiation and evapotranspiration, intercept rainfall, filter air pollutants, store carbon, and shape the thermal and visual quality of streets and public spaces \citep{roy2012benefits,livesley2016urbanforest,nowak2018declining}. These services are inherently spatial: a citywide mean cannot show whether shade is located near homes, active-travel routes, schools, or hot paved surfaces. Urban-forest planning therefore needs maps that resolve neighborhood variation, can be updated when new imagery becomes available, and are explicit about what the mapped objects represent.

Fine-scale canopy mapping remains a measurement challenge. Field inventories provide species, stem diameter, condition, and ownership, while airborne LiDAR resolves height and crown structure. High-resolution optical imagery offers broad, repeatable coverage and can be refreshed when new acquisitions become available, but it represents canopy through spectral and visible-shape evidence rather than direct height. At leaf scale, hyperspectral multi-trait modeling has shown that reflectance can support simultaneous estimation of biochemical and nutritional traits \citep{farajpoor2025multitrait}. Reviews of urban-forest remote sensing therefore emphasize matching each product to its intended use and validating it against an appropriate reference \citep{li2019urbanforestry}. Recent Geocarto International studies follow the same principle by defining a clear measurement problem, quantifying agreement, and connecting the resulting geospatial product to a specific application \citep{jones2025canopyheight,chen2025superres,chen2025unet,chen2025heat}.

Deep learning has expanded the range of vegetation objects that can be mapped from aerial imagery \citep{kattenborn2021cnn}. DeepForest provides transferable crown proposals from RGB imagery \citep{weinstein2019itc,weinstein2020deepforest}, while benchmark studies demonstrate the importance of sensor characteristics, canopy structure, and reference definition \citep{weinstein2021benchmark}. Pixel-segmentation networks can refine boundaries beyond bounding boxes \citep{freudenberg2022itc}, and the Segment Anything Model (SAM) provides promptable masks without local retraining \citep{kirillov2023sam}. Geospatial applications nevertheless show that prompt design, scale, and domain constraints remain decisive \citep{osco2023samrs,wu2023samgeo}. At field scale, UAV multispectral imagery coupled with temporal deep learning has supported early crop-stress detection \citep{narimani2024drone}. Together, these developments support a modular strategy in which a detector proposes candidate objects, spectral information restricts proposals to vegetation, and a promptable model refines visible canopy shape.

Mapping alone, however, does not establish planning relevance. Urban heat reflects the interaction of surface materials, moisture, geometry, and vegetation \citep{oke1982uhi}. Tree cover is often associated with lower daytime land-surface temperature, but the magnitude depends on scale and the surrounding impervious matrix \citep{ziter2019scale,schwaab2021trees}. Canopy is also unequally distributed in many cities. Multi-city studies commonly report less canopy in lower-income or historically disinvested neighborhoods \citep{schwarz2015money,gerrish2018income,hoffman2020historical,locke2021segregation,mcdonald2021disparity}, yet local results can change after accounting for density, land use, and spatial dependence \citep{nesbitt2019access,riley2020equity}. A defensible city study must therefore test rather than presume an equity pattern, distinguish association from causation, and model the fact that neighboring grid cells are not statistically independent.

Davis, California, provides a focused test bed because the city has an adopted Urban Forest Management Plan and a reported 2020 canopy baseline of 26.2\%, while its compact urban form contains mature neighborhoods, newer development, major roads, parks, greenbelts, and an extensive active-travel network \citep{davis2023ufmp,davis2025stateforest}. This study addresses three questions. First, how closely does a DeepForest--NDVI--SAM optical product align, on an identical municipal extent, with a LiDAR-assisted canopy product? Second, how are mapped canopy and crown candidates associated with streets, land-surface temperature, and built form? Third, how do canopy patterns intersect with population, poverty, income, land value, and amenities after spatial dependence is considered, and where do canopy need and heat co-occur? The contribution is an end-to-end measurement-to-screening framework with a shared-extent reference comparison, separate pixel and detection-center metrics, spatial diagnostics, and transparent attention surfaces.

\section{Materials and methods}\label{sec:methods}

\subsection{Study area and common analysis extent}

\begin{figure*}[t]
\centering
\includegraphics[width=\textwidth]{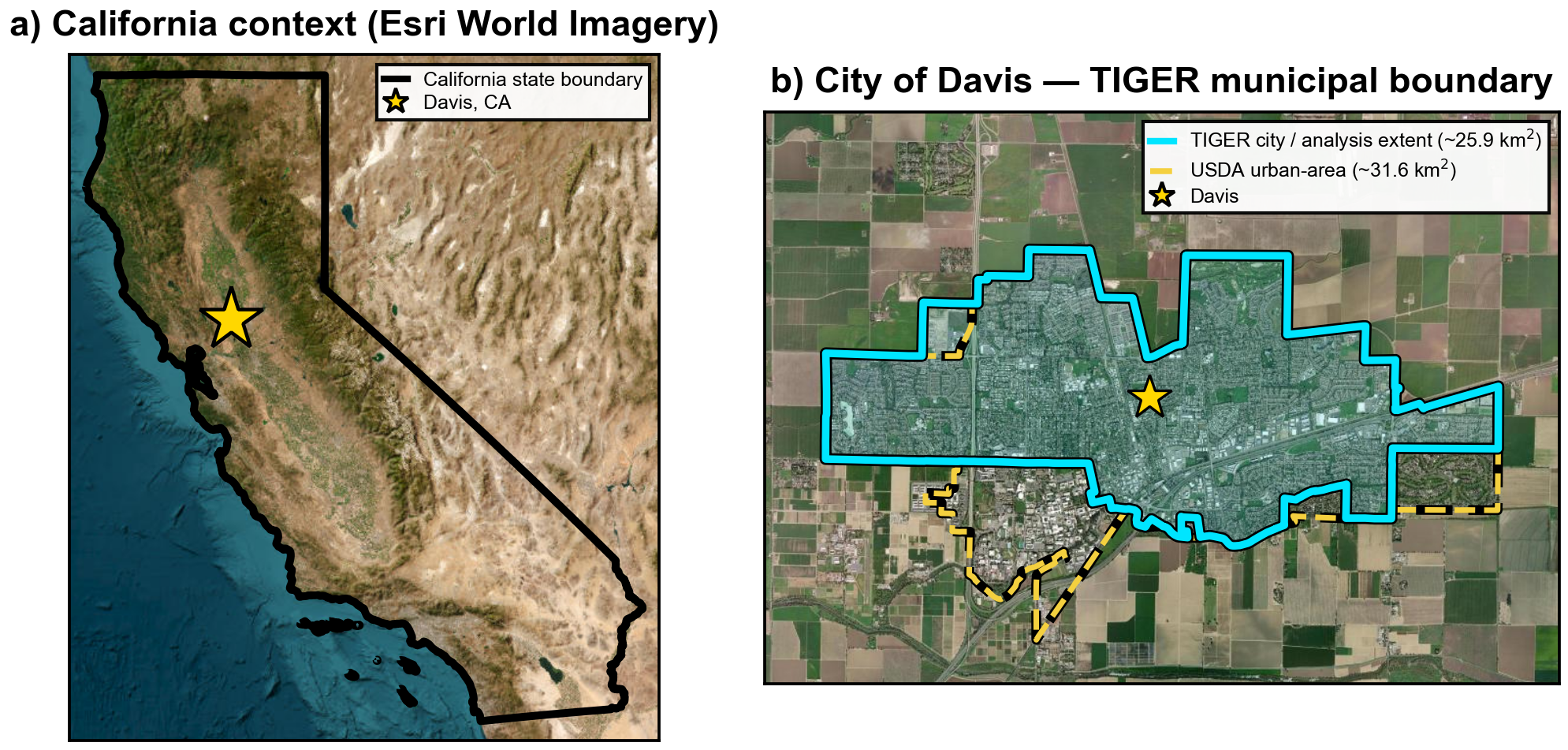}
\caption{Study area and common analysis extent. (a) Davis within California. (b) The 2024 Census TIGER municipal boundary used for all primary analyses (25.92\,km$^2$) and the larger USDA urban-area polygon distributed with the canopy reference. Basemap source: Esri World Imagery (Esri, Maxar, Earthstar Geographics, and the GIS User Community).}
\label{fig:study}
\end{figure*}

The study area was the 2024 Census TIGER/Line Place boundary for Davis, California (GEOID 0618100), covering 25.92\,km$^2$ \citep{census2024tiger}. This municipal polygon was used consistently for mapping, statistical analysis, and external comparison. The USDA canopy package also includes a larger urban-area polygon; Figure~\ref{fig:study} distinguishes the two boundaries and locates Davis within California. All area and distance calculations used UTM Zone 10N, whereas maps are displayed in geographic coordinates. Supplementary Figure~\ref{fig:s1} provides a citywide NAIP overview.

\subsection{Data sources and preprocessing}

Table~\ref{tab:data} identifies every source product, version or vintage, spatial support, provider identifier, and analytical role. The optical base was the 2022 USDA National Agriculture Imagery Program (NAIP) California acquisition, accessed through the Google Earth Engine collection USDA/NAIP/DOQQ. The mosaic provides 0.6\,m red, green, blue, and near-infrared bands and was clipped to the TIGER boundary \citep{usda2022naip,gorelick2017gee}. The comparison layer was the 2022 California Urban Tree Canopy product prepared by EarthDefine and Dewberry for the USDA Forest Service, CAL FIRE, and NOAA. It is a 0.6\,m LiDAR-assisted canopy product for California urban areas \citep{calfire2025canopy}. The reference was used for shared-extent area and pixel agreement; it was not treated as individual-tree ground truth.

Thermal context was obtained from the USGS Landsat 8--9 OLI/TIRS Collection 2 Level-2 surface-temperature product (dataset DOI 10.5066/P9OGBGM6). Cloud, cirrus, shadow, and snow flags were removed, the published scale and offset were applied to ST\_B10, and the clearest scene nearest the NAIP acquisition period was clipped to Davis \citep{eros2020landsat}. The variable is land-surface temperature (LST), not 2\,m air temperature. A coarser reanalysis field varied by only about 0.25\,\textdegree{}C across the city and was excluded from neighborhood inference.

Built and tree-context probabilities came from Dynamic World V1 (GOOGLE/DYNAMICWORLD/V1), a 10\,m Sentinel-2-derived product \citep{brown2022dynamicworld}. Street centerlines, functional classes, bikeways, parcels, parks, greenbelts, and open-space features were obtained from Yolo GIS Cooperative and City of Davis ArcGIS services. Pedestrian ways were retrieved from OpenStreetMap and attributed under the Open Database License \citep{haklay2008osm,osm2026data,yolo2026gis}. Population density and poverty were derived from 2019--2023 American Community Survey five-year estimates and Census geographies; the repository processing produced an income proxy and residential assessed value per acre for contextual analysis \citep{census2023acs}. NLCD 2016 imperviousness was used only as a temporally mismatched sensitivity layer (Supplementary Figure~\ref{fig:s4}) \citep{homer2019nlcd}. The broader Sentinel-2 literature similarly emphasizes matching temporal resolution, spectral information, and model transferability to the operational mapping objective \citep{narimani2026sentinelreview}.

\begin{table*}[t]
\centering
\caption{Data sources, identifiers, analytical roles, and interpretation.}
\label{tab:data}
\footnotesize
\begin{tabular*}{\textwidth}{@{\extracolsep{\fill}}
  >{\raggedright\arraybackslash}p{0.23\textwidth}
  >{\raggedright\arraybackslash}p{0.20\textwidth}
  >{\raggedright\arraybackslash}p{0.23\textwidth}
  >{\raggedright\arraybackslash}p{0.24\textwidth}@{}}
\toprule
\textbf{Product / provider} & \textbf{Vintage and spatial support} & \textbf{Access / identifier} & \textbf{Use and interpretation} \\
\midrule
Census TIGER/Line Place boundary, U.S. Census Bureau
  & 2024 municipal vector; Davis GEOID 0618100; 25.92\,km$^2$
  & Web --- census.gov TIGER/\allowbreak Line 2024
  & Common extent for all primary products and models; not interchangeable with the larger USDA urban-area polygon. \\
NAIP, USDA FSA
  & 2022 California acquisition; 0.6\,m; red, green, blue, NIR
  & GEE --- USDA/\allowbreak NAIP/\allowbreak DOQQ
  & DeepForest input, NDVI vegetation gate, SAM prompts, and optical canopy mapping; no direct height. \\
California Urban Tree Canopy 2022, USDA Forest Service / CAL FIRE / NOAA; EarthDefine and Dewberry
  & 2022; 0.6\,m; LiDAR-assisted
  & Portal --- CAL FIRE Urban Tree Canopy viewer
  & Shared-extent canopy-area and pixel-agreement reference; not individual-tree ground truth. \\
Landsat 8--9 OLI/TIRS Collection 2 Level-2, USGS EROS
  & 2022 scene; ST\_B10; thermal native $\sim$100\,m, delivered at 30\,m
  & DOI --- 10.5066/\allowbreak P9OGBGM6
  & Neighborhood land-surface temperature; LST is not 2\,m air temperature. \\
Dynamic World V1, Google / World Resources Institute
  & 2022 composite; 10\,m class probabilities
  & GEE --- DynamicWorld/\allowbreak V1; DOI 10.1038/\allowbreak s41597-022-01307-4
  & Built probability in the primary model; tree probability in Supplementary Figure~\ref{fig:s5}. \\
American Community Survey 5-year estimates, U.S. Census Bureau
  & 2019--2023 estimates; block-group and related aggregate support
  & API --- census.gov ACS 5-year (2019--2023)
  & Population density, poverty, and income context; estimates retain sampling and aggregation uncertainty. \\
Yolo GIS Cooperative and City of Davis ArcGIS services
  & Vector layers; accessed 2026
  & Service --- ArcGIS Online (Yolo/\allowbreak Davis)
  & Road centerlines/classes, bikeways, parcels, parks, greenbelts, and open space; mixed source vintages. \\
OpenStreetMap contributors
  & Pedestrian and footway vectors; accessed 2026
  & Web --- openstreetmap.org (ODbL)
  & Pedestrian-network context; coverage and tagging completeness vary. \\
NLCD 2016, USGS
  & 2016; 30\,m percent impervious
  & DOI --- 10.5066/\allowbreak P937PN4Z
  & Sensitivity and visual context only because of temporal mismatch. \\
\bottomrule
\end{tabular*}
\end{table*}

\subsection{Candidate crown detection and canopy delineation}

The implemented workflow is summarized in Table~\ref{tab:pipeline}. DeepForest 1.3.3 with pretrained NEON.pt weights generated candidate crown boxes from overlapping 512\,$\times$\,512 RGB patches with 128-pixel overlap and a score threshold of 0.05. For each proposal, mean normalized difference vegetation index (NDVI) was computed from the NAIP near-infrared (NIR) and red (R) bands:
\begin{equation}
\mathrm{NDVI} = \frac{\mathrm{NIR} - \mathrm{R}}{\mathrm{NIR} + \mathrm{R}}
\end{equation}
Here NIR and R are the pixel values in the NAIP near-infrared and red bands \citep{tucker1979ndvi}. Proposals with mean box NDVI below 0.20 were removed. Non-maximum suppression with an intersection-over-union threshold of 0.30 consolidated duplicate proposals from overlapping detector patches. Retained boxes were clipped to the TIGER municipal boundary and interpreted as candidate crowns.

Each retained box was passed to SAM ViT-B as an RGB box prompt within 800-pixel windows with 100-pixel overlap. Near-infrared information did not enter SAM directly. Instead, the local canopy mask was defined as the union of the SAM prediction and pixels satisfying NDVI $\geq$ 0.20 inside the corresponding proposal box. This crown-anchored rule uses object location, spectral vegetation evidence, and visible shape while limiting expansion into vegetation unrelated to a proposal. A related NAIP workflow has shown that promptable SAM refinement can complement learned segmentation where visible agricultural boundaries are incomplete or fragmented \citep{narimani2026farmland}.

The crown-anchored base mask recovered only a subset of reference canopy area because much structural canopy lies outside detected boxes. A second stage therefore unioned the base mask with a spectral--texture infill: pixels satisfying NDVI $\geq$ 0.18 and local NIR standard deviation $\geq$ 3 (7-pixel window), followed by 3$\times$3 binary closing and removal of connected components smaller than 25\,m$^2$. Thresholds were tuned on the western half of the city and evaluated on the eastern half against the USDA/CAL FIRE reference (holdout IoU 0.707 west, 0.690 east), reducing the risk of tuning on the evaluation product itself. SAM re-prompting of infill patches was tested and did not improve agreement; the transparent spectral--texture rule was retained.

\begin{table*}[t]
\centering
\caption{Implemented settings for the optical mapping workflow.}
\label{tab:pipeline}
\begin{tabular*}{\textwidth}{@{\extracolsep{\fill}}p{0.16\textwidth}p{0.42\textwidth}p{0.34\textwidth}@{}}
\toprule
\textbf{Stage} & \textbf{Implementation / setting} & \textbf{Analytical meaning} \\
\midrule
Imagery & NAIP 2022; 0.6\,m; R, G, B, NIR & Four-band optical source. \\
DeepForest & Version 1.3.3; NEON.pt; 512\,px patches; 128\,px overlap; score $\geq$ 0.05 & Broad candidate generation followed by spectral and spatial filtering. \\
Vegetation gate & Mean proposal-box NDVI $\geq$ 0.20 & NIR enters through NDVI, not through SAM. \\
Duplicate consolidation & Non-maximum suppression; IoU threshold 0.30 & Reduces duplicated proposals across overlapping detector patches. \\
SAM & ViT-B; RGB box prompts; 800\,px windows; 100\,px overlap & Prompted refinement of visible canopy shape. \\
Canopy rule & SAM mask $\cup$ pixels with NDVI $\geq$ 0.20 inside the retained proposal box & Combines visible shape and vegetation evidence while remaining crown anchored. \\
Window merge & Sequential writes in the frozen run & Crown-anchored base mask before canopy expansion. \\
Canopy expansion & NDVI $\geq$ 0.18 and local NIR texture ($\geq$ 3 DN in 7\,px window); 3$\times$3 closing; patches $<$ 25\,m$^2$ removed & Unions crown-anchored mask with spectral--texture infill; thresholds tuned on the west half and evaluated on the east half. \\
Spatial summary & 100\,m grid in UTM Zone 10N, clipped to TIGER & Partial boundary cells retain their true area. \\
\bottomrule
\end{tabular*}
\end{table*}

\subsection{Fair external comparison}

The optical canopy mask and the USDA/CAL FIRE product were clipped to the identical TIGER boundary and aligned to a common grid using nearest-neighbor resampling. Agreement was summarized by mapped area, reference area, intersection, reference-only area, optical-only area, area recovery, pixel precision, pixel recall, intersection over union (IoU), and Dice coefficient. Detection-center agreement was computed separately as the proportion of optical candidate centers falling on reference canopy. For binary comparison with true positives $TP$ (canopy mapped by both products), false positives $FP$ (optical-only canopy), and false negatives $FN$ (reference-only canopy), the overlap metrics were
\begin{equation}
\mathrm{IoU} = \frac{TP}{TP + FP + FN}
\end{equation}
and
\begin{equation}
\mathrm{Dice} = \frac{2\,TP}{2\,TP + FP + FN}
\end{equation}
These are pixel-level comparison terms; they do not represent individual-tree matches. Reporting area and pixel metrics separately follows established remote-sensing accuracy guidance and preserves the distinction between a crown-candidate product and a structural canopy layer \citep{olofsson2014goodpractices,weinstein2021benchmark}.

\subsection{Municipal grid and contextual variables}

A 100\,m grid in UTM Zone 10N was clipped to the municipal boundary. For each cell, we calculated candidate-crown count, effective area, candidates per hectare, optical canopy percentage, mean Landsat LST, Dynamic World built probability, road and pedestrian-way footprint, and the contextual variables in Table~\ref{tab:data}. Point and raster layers were summarized by spatial join or zonal statistics; polygon attributes were transferred using the repository enrichment workflow. Partial boundary cells retained their true area. The resulting grid included 2{,}747 descriptive cells, with complete-case subsets reported for each statistical model.

\subsection{Association tests and spatial models}

Pairwise associations with canopy percentage were screened using Spearman rank correlation. P values for the screened variables were adjusted with the Benjamini--Hochberg false-discovery-rate procedure \citep{benjamini1995fdr}. The canopy--LST relationship was also evaluated with partial Spearman correlation controlling for Dynamic World built probability. Kruskal--Wallis tests compared canopy distributions across poverty and income quintiles; epsilon-squared quantified effect size.

Global Moran's $I$ based on eight-nearest-neighbor spatial weights (KNN-8) quantified clustering in canopy, LST, built probability, poverty, and model residuals \citep{anselin1995lisa}. A non-spatial ordinary least-squares model used canopy percentage as the outcome and LST, Dynamic World built probability, population density, and poverty as predictors. Residual spatial autocorrelation motivated generalized method-of-moments spatial-lag and spatial-error models. The spatial-lag specification was
\begin{equation}
y = \rho\,W y + X\beta + \varepsilon
\end{equation}
where $y$ is the $n \times 1$ vector of canopy percentages, $W$ is the row-standardized KNN-8 spatial-weights matrix, $\rho$ is the spatial autoregressive coefficient, $X$ is the predictor matrix, $\beta$ is the coefficient vector, and $\varepsilon$ is the residual vector. Model choice considered fit, residual diagnostics, coefficient stability, and variance-inflation factors rather than relying on a single goodness-of-fit statistic \citep{dormann2007spatial}.

\subsection{Transparent attention screening}

Two continuous attention surfaces translated the analysis into reproducible planning screens. The first combined only canopy need and heat. For cell $i$, percentile ranks were oriented so that lower canopy and higher LST both increased attention:
\begin{equation}
A_i^{\mathrm{shade}} = \tfrac{1}{2}\left[1 - P_i(C_i)\right] + \tfrac{1}{2}\,P_i(T_i)
\end{equation}
where $A_i^{\mathrm{shade}}$ is the shade-attention score for cell $i$; $P_i(\cdot)$ is the empirical percentile-rank operator scaled from 0 to 1; $C_i$ is canopy percentage; and $T_i$ is Landsat LST. For categorical summaries, low canopy was defined as $C_i \leq 19.6\%$ (25th percentile), hot as $T_i \geq 33.84\,$\textdegree{}C (75th percentile), and higher poverty as $\geq 33.38\%$ among cells with poverty estimates.

The second surface combined nine disclosed factors. Direction was specified before scoring: lower canopy, lower residential value per acre, higher LST, higher built probability, higher population density, higher poverty, lower income, greater distance from parks, and greater distance from bikeways. Each weight was proportional to the absolute Spearman association with canopy:
\begin{equation}
w_j = \frac{\left|\rho_{jC}\right|}{\sum_{k=1}^{K}\left|\rho_{kC}\right|}
\end{equation}
\begin{equation}
A_i^{\mathrm{multi}} = \sum_{j=1}^{K} w_j\,P_i\!\left(d_j\,x_{ij}\right)
\end{equation}
where $\rho_{jC}$ is the Spearman coefficient between factor $j$ and canopy, $K = 9$ is the number of included factors, $w_j$ is the normalized non-negative weight, $x_{ij}$ is factor $j$ in cell $i$, and $d_j \in \{-1,+1\}$ orients the factor so that larger values represent greater attention. Tree count was omitted because it was redundant with canopy, while road and pedestrian-way footprints were omitted because their absolute correlations were near zero. The result is a transparent sensitivity scenario, not an official equity index \citep{malczewski2006gis}.

\section{Results}\label{sec:results}

\subsection{Crown-anchored canopy mapping}

\begin{figure*}[t]
\centering
\includegraphics[width=\textwidth]{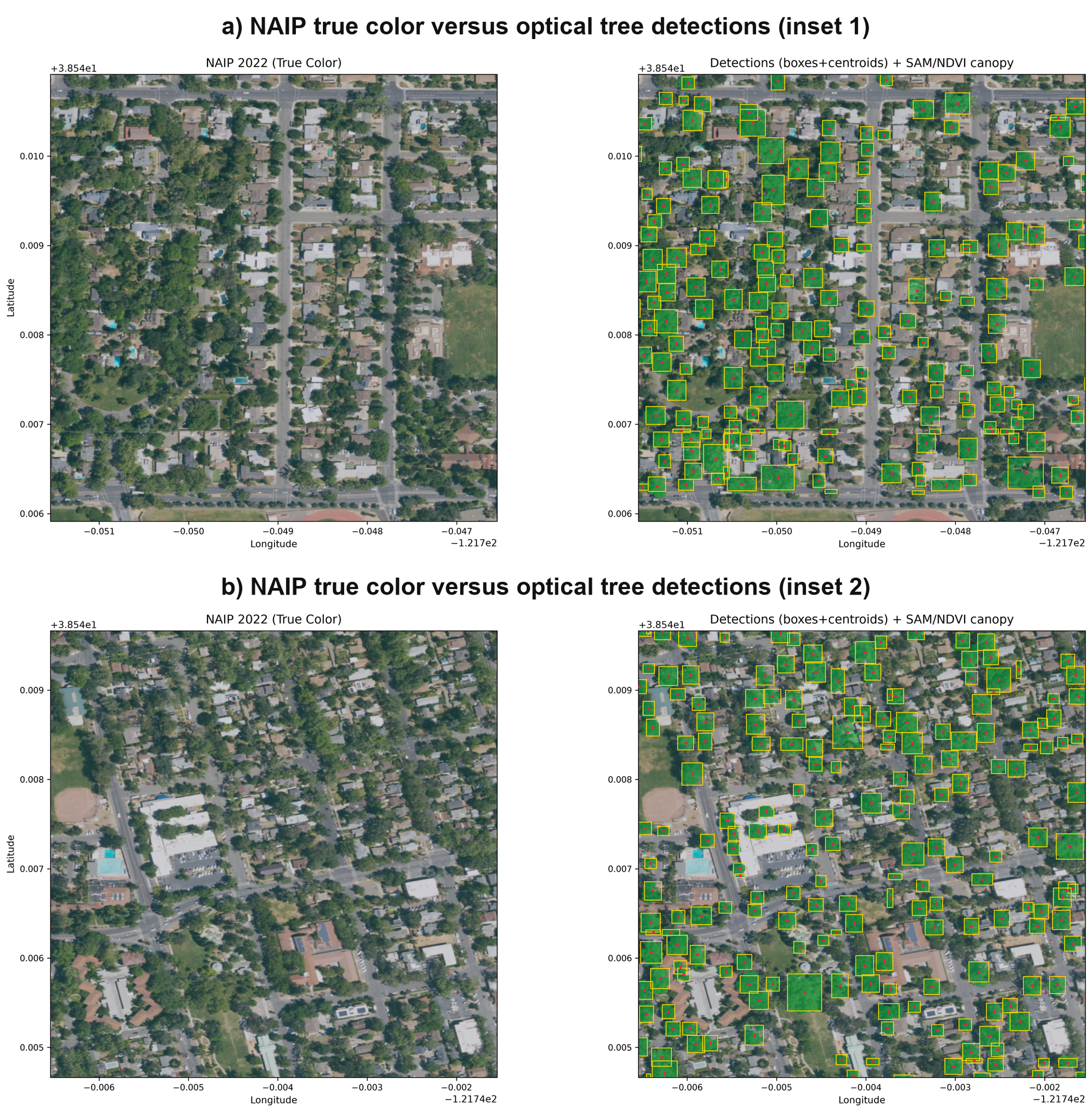}
\caption{Representative optical mapping examples. Panels (a) and (b) compare 2022 NAIP true color with DeepForest candidate boxes (yellow), candidate centers (red), and the SAM/within-box NDVI canopy mask (green) for two neighborhoods. The examples show the crown-anchored behavior of the mapping chain.}
\label{fig:examples}
\end{figure*}

The workflow retained 11{,}741 candidate crowns after NDVI filtering, non-maximum suppression, and municipal clipping (Table~\ref{tab:descriptive}). Figure~\ref{fig:examples} presents two representative neighborhoods in which proposal boxes and centroids are overlaid with the SAM/NDVI canopy. The mapped candidates correspond closely to visually distinct tree crowns across residential streets, institutional parcels, and mixed urban vegetation, illustrating the intended object-anchored behavior of the workflow.

\begin{figure*}[t]
\centering
\includegraphics[width=\textwidth]{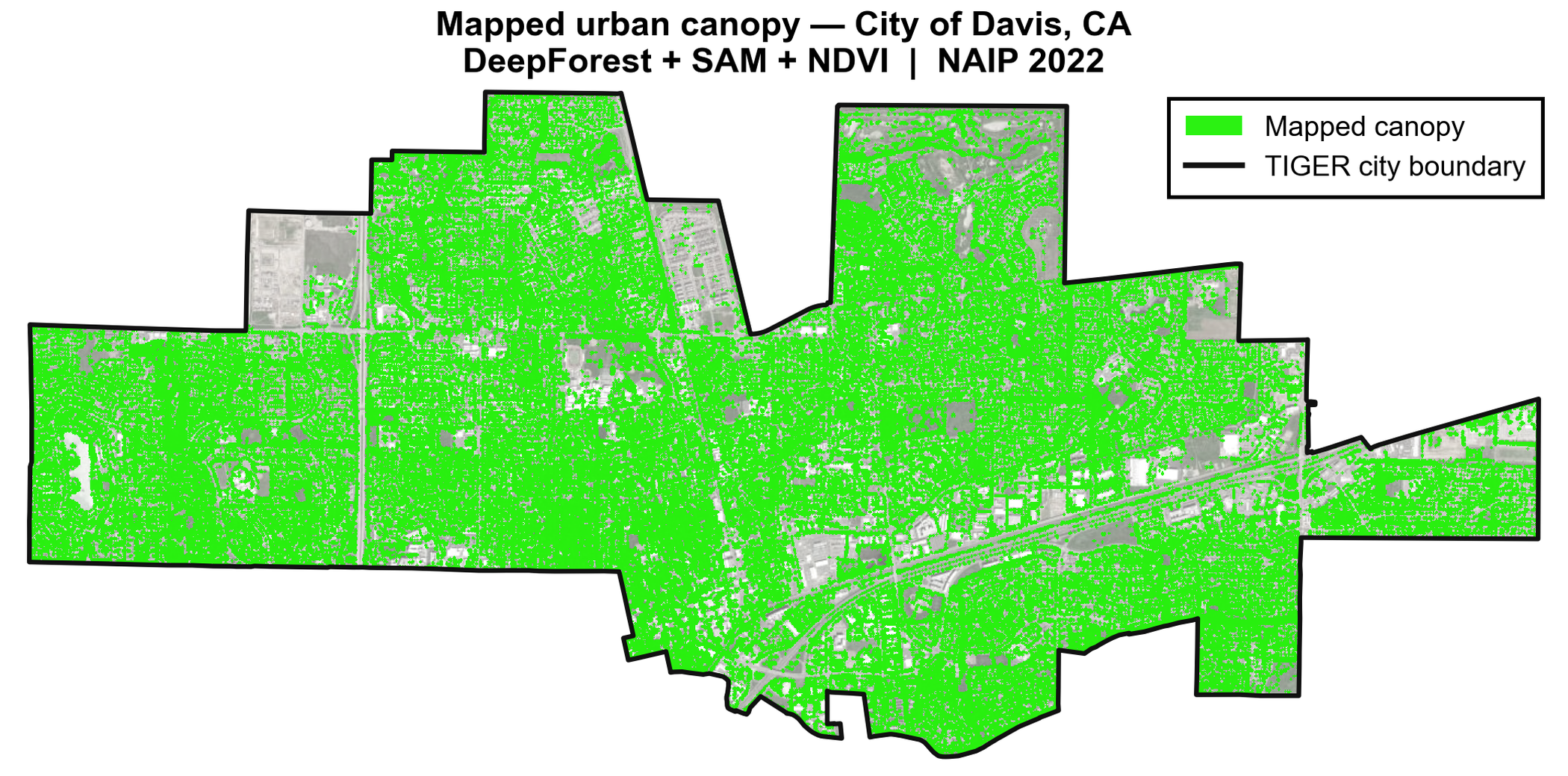}
\caption{Citywide optical canopy product within the Census TIGER boundary. Green pixels show the crown-anchored SAM/NDVI mask unioned with NDVI--texture infill after DeepForest proposal filtering and non-maximum suppression. The mapped area is 7.71\,km$^2$, or 29.8\% of the municipal extent.}
\label{fig:canopy}
\end{figure*}

Across the TIGER municipal boundary, the optical product mapped 7.71\,km$^2$ of canopy, equivalent to 29.8\% of city area (Figure~\ref{fig:canopy}). The 2{,}747-cell descriptive grid had mean canopy of 29.6\% (SD 15.3\%; median 30.5\%; maximum 100\%) and a mean of 4.22 candidate crowns per cell (SD 3.35; median 4; maximum 26). Mean candidate density was 4.42\,ha$^{-1}$. These outputs provide a high-resolution, updateable optical layer for subsequent spatial analysis.

\begin{table*}[t]
\centering
\caption{Descriptive mapping and infrastructure results.}
\label{tab:descriptive}
\begin{tabular*}{\textwidth}{@{\extracolsep{\fill}}p{0.30\textwidth}p{0.24\textwidth}p{0.38\textwidth}@{}}
\toprule
\textbf{Metric} & \textbf{Value} & \textbf{Interpretation} \\
\midrule
Retained crown candidates & 11{,}741 & After NDVI filtering, NMS, and municipal clipping. \\
Candidates in fair comparison clip & 11{,}667 & Used for detection-center agreement. \\
Analysis cells & 2{,}747 & Boundary-clipped 100\,m descriptive grid. \\
Optical canopy area & 7.71\,km$^2$ & 29.8\% of the 25.92\,km$^2$ city. \\
Mean canopy per cell & 29.6\% (SD 15.3; median 30.5; max 100) & Optical canopy percentage. \\
Mean candidates per cell & 4.27 (SD 3.35; median 4; max 26) & Candidate-crown count. \\
Mean candidate density & 4.42\,ha$^{-1}$ & Area adjusted for partial cells. \\
Mean road footprint & 9.48\% & Per grid cell. \\
Mean pedestrian-way footprint & 2.32\% & Per grid cell. \\
Candidates within 15\,m of road & $\sim$49\%; median distance 16\,m & Street-adjacent component of the mapped urban forest. \\
\bottomrule
\end{tabular*}
\end{table*}

\subsection{Shared-extent agreement with the LiDAR-assisted canopy product}

\begin{figure*}[t]
\centering
\includegraphics[width=\textwidth]{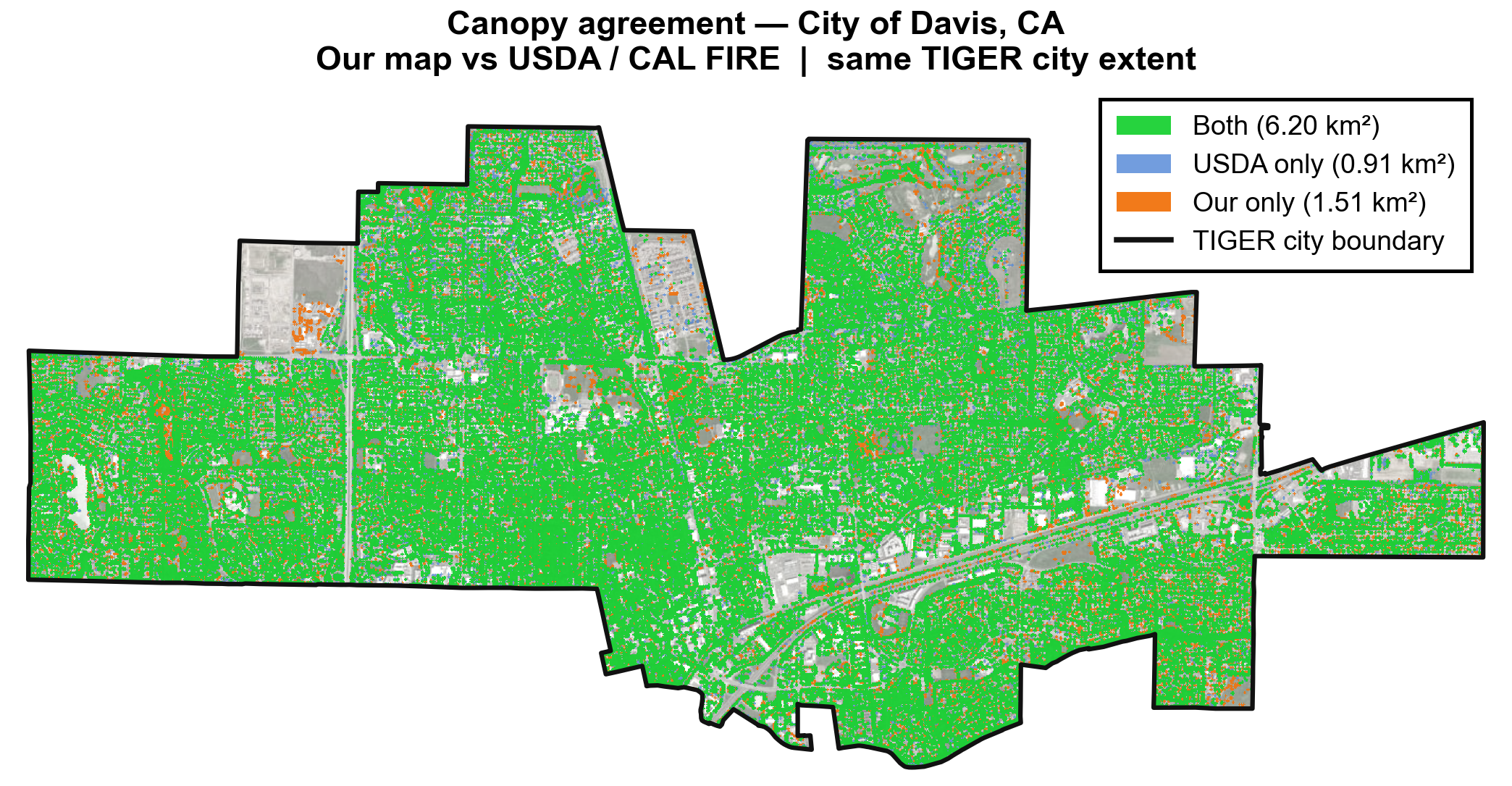}
\caption{Shared-extent agreement between the optical canopy product and the 2022 USDA/CAL FIRE LiDAR-assisted canopy product. Green indicates agreement (6.20\,km$^2$), blue indicates reference-only canopy (0.91\,km$^2$), and orange indicates optical-only canopy (1.51\,km$^2$). Both layers were clipped to the same TIGER boundary.}
\label{fig:fair}
\end{figure*}

The common-extent comparison demonstrates strong locational consistency where the optical workflow identified canopy (Figure~\ref{fig:fair}; Table~\ref{tab:fair}). Of the 7.71\,km$^2$ mapped optically, 6.20\,km$^2$ coincided with the LiDAR-assisted product and 1.51\,km$^2$ was optical-only, yielding pixel precision of 0.804 and Dice of 0.837. The structural product represented 7.11\,km$^2$ of canopy on the same city boundary; the optical layer recovered 108.5\% of that extent, with IoU 0.719 and pixel recall 0.873. The products are now near parity in total canopy area while retaining distinct sensing bases: LiDAR-assisted structure versus reproducible NAIP optical inference anchored on detected crowns and NDVI--texture infill.

The fair city clip contained 11{,}667 candidate detections. Of these, 11{,}366 centroids fell on reference canopy and 301 did not, corresponding to detection-center precision of 0.974. Thus, candidate locations were highly consistent with the structural canopy layer even though the optical surface represented a subset of total canopy. Pixel-area and center-location metrics are reported together because they describe different, complementary properties of the mapping system. Supplementary Figure~\ref{fig:s2} shows the corresponding detection-center overlay.

\begin{table*}[t]
\centering
\caption{Shared-extent comparison with the 2022 USDA/CAL FIRE LiDAR-assisted canopy product.}
\label{tab:fair}
\begin{tabular*}{\textwidth}{@{\extracolsep{\fill}}p{0.30\textwidth}p{0.24\textwidth}p{0.38\textwidth}@{}}
\toprule
\textbf{Quantity} & \textbf{Value} & \textbf{Interpretation} \\
\midrule
Municipal comparison area & 25.92\,km$^2$ & Census TIGER Davis city. \\
Optical canopy area & 7.71\,km$^2$ (29.8\%) & Expanded optical product (crown anchor + NDVI--texture infill). \\
LiDAR-assisted reference canopy & 7.11\,km$^2$ (27.4\%) & Structural canopy product on the same extent. \\
Intersection & 6.20\,km$^2$ & Mapped by both products. \\
Reference-only area & 0.91\,km$^2$ & Additional canopy represented by the structural product. \\
Optical-only area & 1.51\,km$^2$ & Mapped optically but not in the reference. \\
Reference-area representation & 108.5\% & Optical canopy area / reference canopy area. \\
Pixel IoU & 0.719 & Intersection relative to union. \\
Pixel Dice & 0.837 & Overlap similarity. \\
Pixel precision & 0.804 & Most optical canopy pixels coincide with reference canopy. \\
Pixel recall & 0.873 & Optical product captures most reference canopy on the shared extent. \\
Candidate centers on reference canopy & 11{,}366 / 11{,}667 (0.974) & High detection-center location agreement; not object recall. \\
\bottomrule
\end{tabular*}
\end{table*}

\subsection{Street and active-travel context}

\begin{figure*}[t]
\centering
\includegraphics[width=\textwidth]{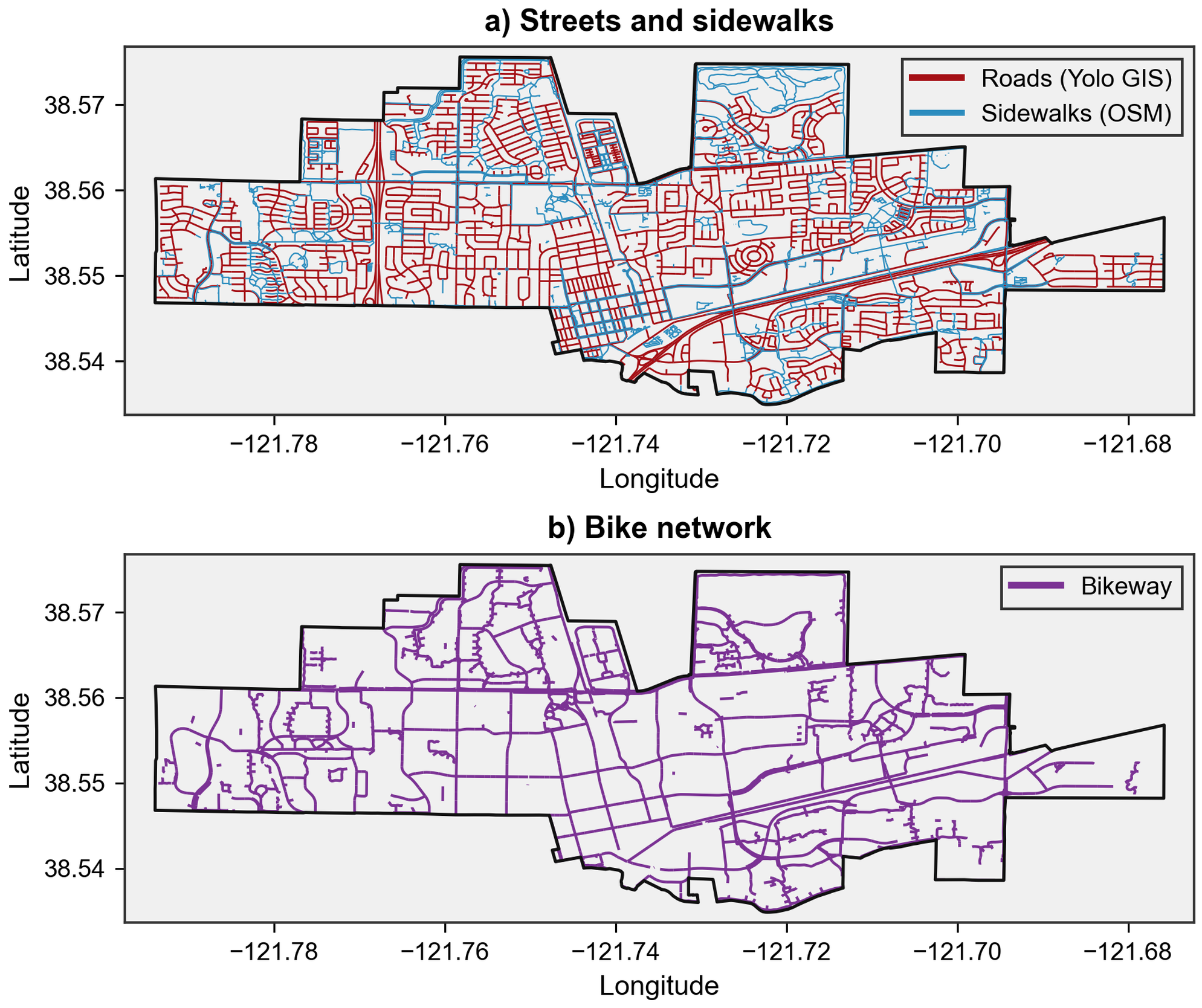}
\caption{Transportation context. (a) Yolo GIS road centerlines and OpenStreetMap pedestrian ways. (b) Bikeway network. Approximately 49\% of retained candidates are within 15\,m of a road (median 16\,m). Map data \copyright{} OpenStreetMap contributors, ODbL, for pedestrian ways.}
\label{fig:transport}
\end{figure*}

Roads occupied a mean 9.48\% of each 100\,m cell and pedestrian ways 2.32\%. Approximately 49\% of retained candidates were within 15\,m of a road, with a median road distance of 16\,m. Figure~\ref{fig:transport}a shows the Yolo GIS road network and OpenStreetMap pedestrian ways, while Figure~\ref{fig:transport}b shows the city bikeway network; road functional classes are shown in Supplementary Figure~\ref{fig:s3}. The near-road concentration indicates a pronounced street-tree component; by contrast, road-footprint percentage itself had little bivariate association with canopy, showing that proximity and network geometry carry different information from area fraction.

\subsection{Canopy, surface temperature, and built form}

\begin{figure*}[t]
\centering
\includegraphics[width=\textwidth]{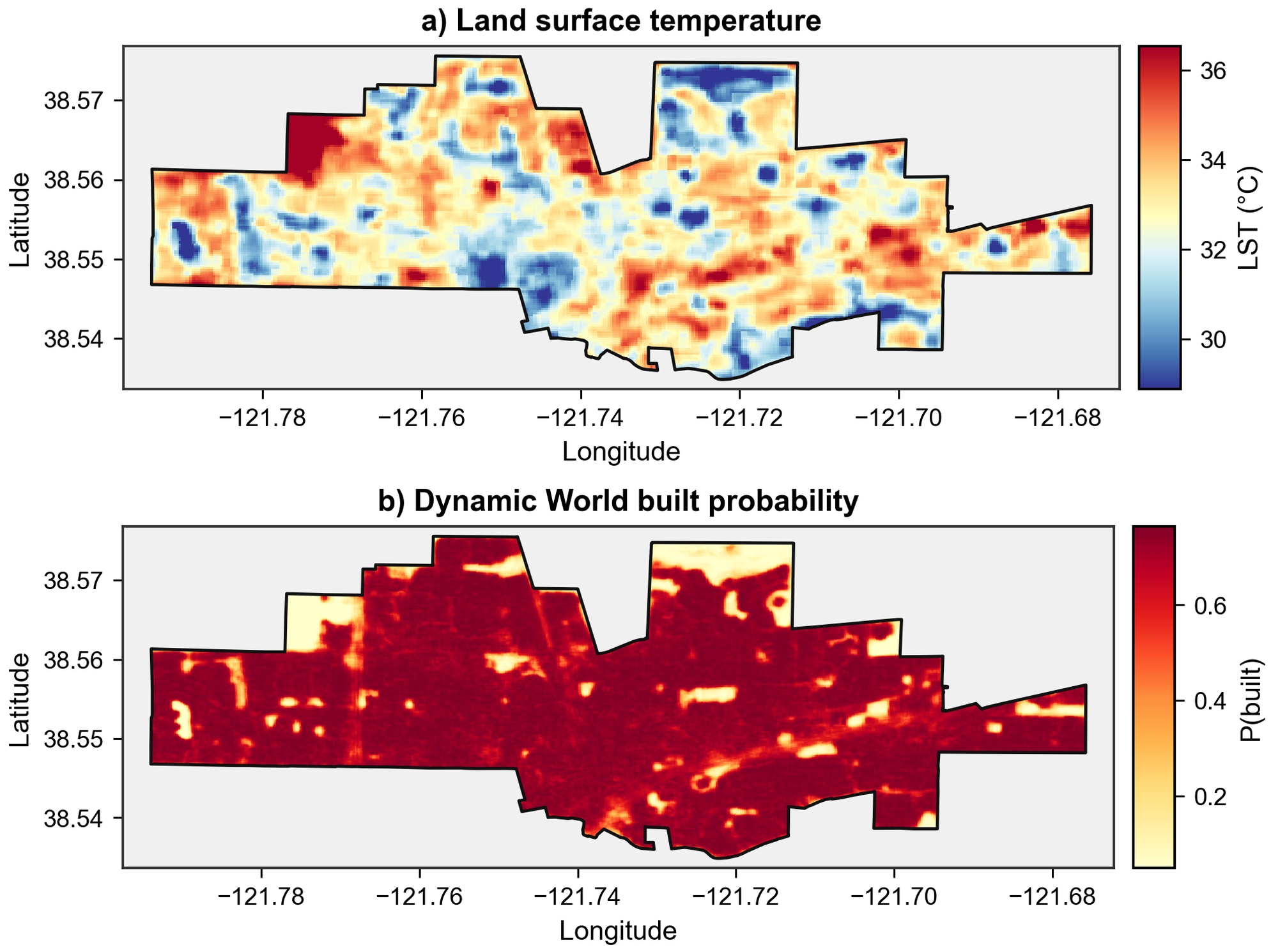}
\caption{Thermal and built-form context. (a) Landsat Collection 2 land-surface temperature; values are surface temperature, not 2\,m air temperature. (b) Dynamic World built probability. Canopy percentage was inversely associated with LST after FDR correction and spatial adjustment. Dynamic World data: Google and World Resources Institute, CC BY 4.0.}
\label{fig:thermal}
\end{figure*}

Landsat LST and Dynamic World built probability exhibited clear neighborhood structure (Figure~\ref{fig:thermal}). Canopy percentage was inversely associated with LST (Spearman $\rho = -0.477$, FDR $q = 8.1 \times 10^{-153}$). The partial association controlling for built probability was $\rho = -0.551$ ($p = 3.2 \times 10^{-216}$; $n = 2{,}719$), indicating that the canopy--temperature relationship remained substantial after accounting for built context. Dynamic World built probability was positively associated with canopy in the bivariate screen ($\rho = +0.299$), underscoring the mixed land-use character of Davis and the need for spatial adjustment.

\subsection{Social and amenity context is locally contingent}

\begin{figure*}[t]
\centering
\includegraphics[width=\textwidth]{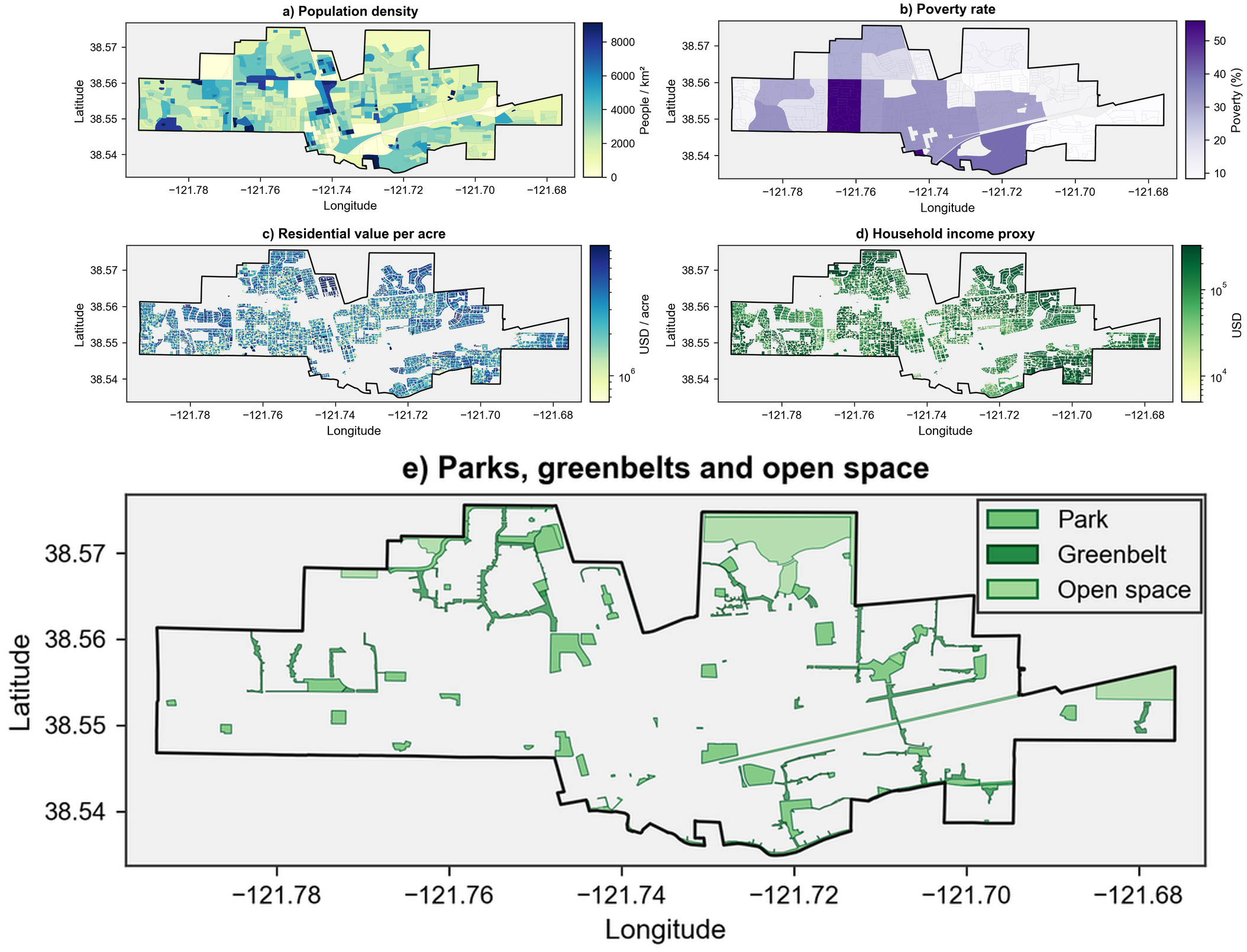}
\caption{Social, land-market, and amenity context: (a) population density, (b) poverty rate, (c) residential assessed value per acre, (d) household income proxy, and (e) parks, greenbelts, and open space. Variables differ in spatial support and are interpreted as contextual indicators.}
\label{fig:social}
\end{figure*}

The contextual maps reveal substantial within-city variation (Figure~\ref{fig:social}). Bivariate canopy associations were negative for residential value per acre ($\rho = -0.277$), household income proxy ($\rho = -0.076$), and distance to parks ($\rho = -0.176$), and positive for population density ($\rho = +0.351$) and poverty ($\rho = +0.185$); all were FDR-significant (Table~\ref{tab:spearman}). These signs do not form a simple affluence gradient. They reflect the local arrangement of mature neighborhoods, student-oriented housing, institutional land, greenbelts, and urban-edge development, and therefore require interpretation alongside the spatial model rather than as standalone equity effects.

Canopy distributions differed across poverty quintiles (Kruskal--Wallis $H = 155.91$, $p = 1.10 \times 10^{-32}$, $\varepsilon^2 = 0.061$, $n = 2{,}481$) and income quintiles ($H = 14.31$, $p = 0.0064$, $\varepsilon^2 = 0.006$, $n = 1{,}639$). The larger poverty effect size indicates stronger distributional separation across poverty groups, although the direction and meaning of that separation remain local and non-causal.

\begin{table*}[t]
\centering
\caption{FDR-adjusted bivariate Spearman associations with optical canopy percentage.}
\label{tab:spearman}
\begin{tabular*}{\textwidth}{@{\extracolsep{\fill}}p{0.34\textwidth}rrl@{}}
\toprule
\textbf{Variable} & \textbf{Spearman $\rho$} & \textbf{FDR $q$} & \textbf{Inference} \\
\midrule
Candidate count & $+0.689$ & $<0.001$ & *** \\
Residential value per acre & $-0.277$ & $2.0 \times 10^{-31}$ & *** \\
Land-surface temperature & $-0.477$ & $8.1 \times 10^{-153}$ & *** \\
Dynamic World built probability & $+0.299$ & $1.0 \times 10^{-56}$ & *** \\
Population density & $+0.351$ & $3.4 \times 10^{-79}$ & *** \\
Poverty rate & $+0.185$ & $5.0 \times 10^{-20}$ & *** \\
Household income proxy & $-0.076$ & $0.003$ & ** \\
Distance to park & $-0.176$ & $4.9 \times 10^{-20}$ & *** \\
Distance to bikeway & $-0.102$ & $1.7 \times 10^{-7}$ & *** \\
Pedestrian-way footprint & $+0.012$ & $0.63$ & ns \\
Road footprint & $-0.004$ & $0.90$ & ns \\
\bottomrule
\end{tabular*}
\begin{minipage}{\textwidth}
\vspace{2pt}
\footnotesize Notes: Bivariate screen on the 100\,m grid; $q$ values use Benjamini--Hochberg correction. *** $q < 0.001$; ** $q < 0.01$; ns, not significant.
\end{minipage}
\end{table*}

\subsection{Spatial dependence changes the interpretation}

\begin{figure*}[t]
\centering
\includegraphics[width=\textwidth]{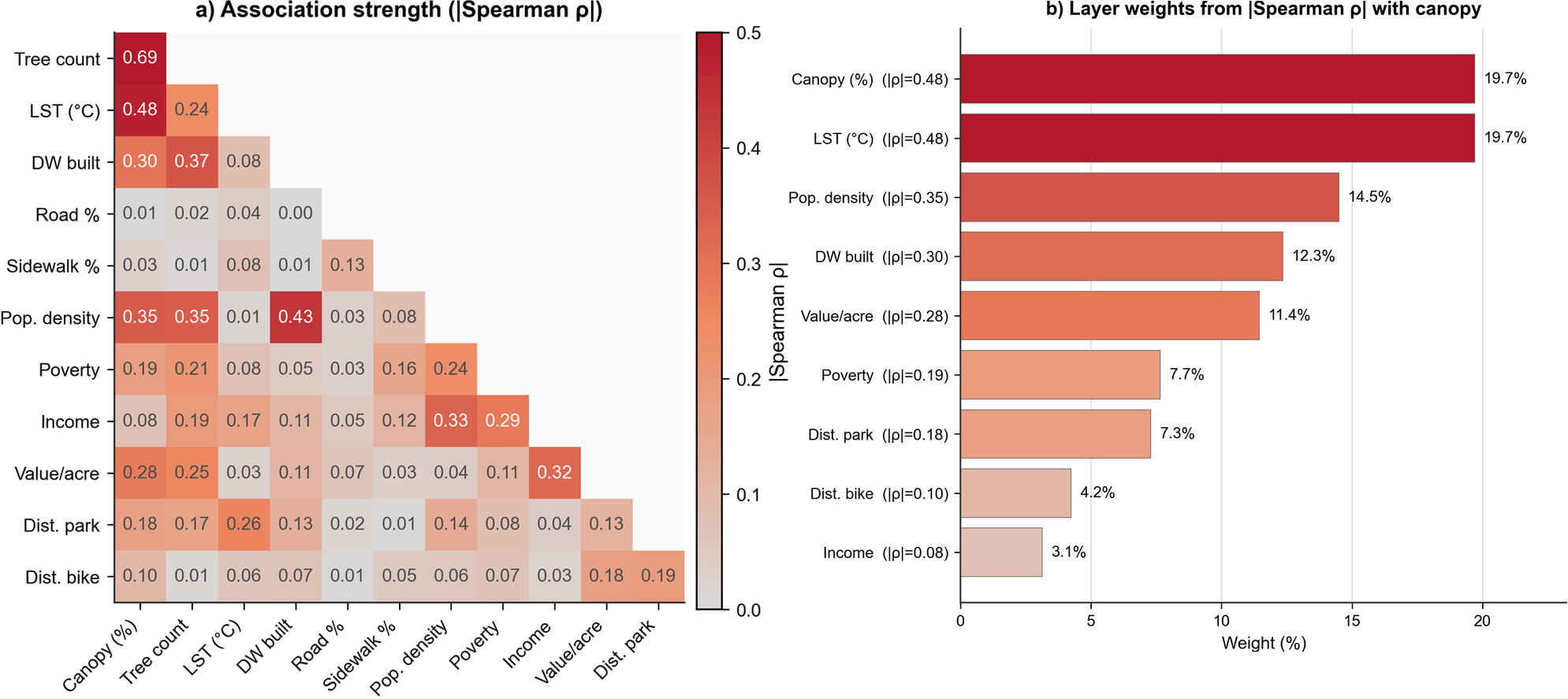}
\caption{Association screen and weight derivation. (a) Lower-triangle matrix of absolute Spearman correlations among canopy and screened layers; FDR-adjusted results for canopy are reported in Table~\ref{tab:spearman}. (b) Normalized multi-factor weights proportional to absolute Spearman association with canopy.}
\label{fig:assoc}
\end{figure*}

Canopy and contextual indicators were spatially clustered. Global Moran's $I$ was approximately 0.58 for canopy, 0.68 for LST, 0.63 for Dynamic World built probability, and 0.93 for poverty; permutation $p$ values were 0.001. The OLS model for 2{,}459 complete core cases explained 38.4\% of canopy variation, but residual Moran's $I$ remained 0.338 ($p = 0.001$). Figure~\ref{fig:assoc} summarizes the pairwise association structure and the transparent derivation of the screening weights.

The spatial-lag model was therefore preferred over OLS and the spatial-error alternative (Table~\ref{tab:spatial}). Its pseudo-$R^2$ was approximately 0.559 and the neighboring-canopy coefficient was $\rho = 0.513$ ($p = 2.8 \times 10^{-30}$). Conditional on the other predictors and spatial lag, a 1\,\textdegree{}C higher LST was associated with 3.00 percentage points lower canopy ($p = 3.6 \times 10^{-51}$), Dynamic World built probability had a positive coefficient of 20.0 ($p = 1.6 \times 10^{-29}$), population density was not significant ($p = 0.65$), and poverty retained a small positive coefficient of 0.0430 ($p = 0.0072$). Core-predictor VIF values of 1.08--1.15 indicated little multicollinearity.

\begin{table*}[t]
\centering
\caption{Spatial diagnostics and preferred canopy model.}
\label{tab:spatial}
\begin{tabular*}{\textwidth}{@{\extracolsep{\fill}}p{0.28\textwidth}p{0.28\textwidth}rp{0.24\textwidth}@{}}
\toprule
\textbf{Model / term} & \textbf{Estimate or fit} & \textbf{$p$ value} & \textbf{Interpretation} \\
\midrule
Global Moran's $I$: canopy & $\sim$0.58 & 0.001 & Spatial clustering. \\
OLS fit & $R^2 = 0.384$; adj. $R^2 \approx 0.383$; AIC $\approx$ 18{,}968 & --- & Non-spatial baseline. \\
OLS residual Moran's $I$ & 0.338 & 0.001 & Residual dependence supports spatial modeling. \\
Spatial-lag pseudo-$R^2$ & $\sim$0.559 & --- & Preferred fit. \\
Spatial lag of canopy ($\rho$) & 0.513 & $2.8 \times 10^{-30}$ & Strong neighborhood structure. \\
LST coefficient & $-3.00$ & $3.6 \times 10^{-51}$ & Higher LST associated with lower canopy. \\
Dynamic World built coefficient & $+20.0$ & $1.6 \times 10^{-29}$ & Positive conditional association. \\
Population-density coefficient & $\sim$0 & 0.65 & No adjusted association. \\
Poverty coefficient & $+0.0430$ & 0.0072 & Small positive conditional association. \\
VIF range & 1.08--1.15 & --- & No serious multicollinearity. \\
\bottomrule
\end{tabular*}
\begin{minipage}{\textwidth}
\vspace{2pt}
\footnotesize Notes: Core complete-case sample $n = 2{,}459$. Spatial weights are row-standardized KNN-8. Coefficients are associations, not causal effects.
\end{minipage}
\end{table*}

\subsection{Two transparent screening surfaces}

\begin{figure*}[t]
\centering
\includegraphics[width=\textwidth]{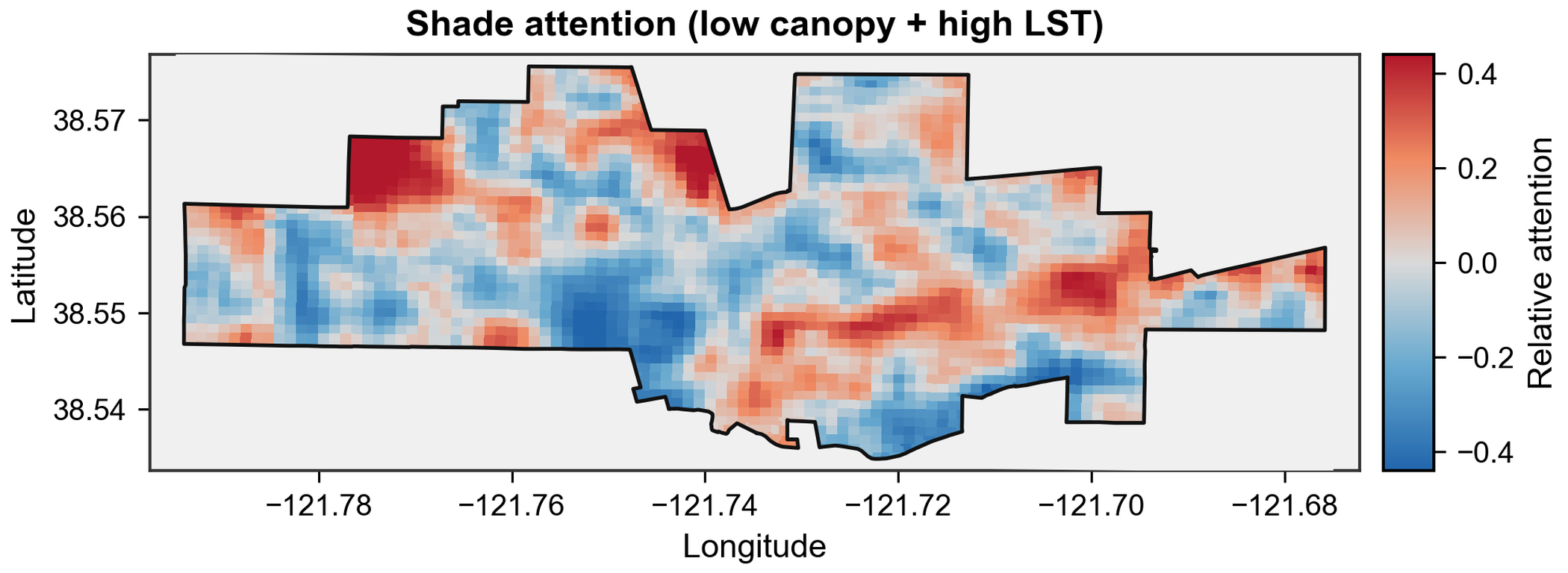}
\caption{Shade-attention surface based on equal-weight low canopy and high Landsat LST. Positive values indicate stronger co-location of canopy need and heat. The surface is a transparent screening layer rather than a causal heat-exposure model.}
\label{fig:shade}
\end{figure*}

The shade-attention surface in Figure~\ref{fig:shade} combines canopy deficit and LST with equal weight. Of the classified cells, 1{,}689 (62.1\%) were in neither extreme, 349 (12.8\%) were low-canopy only, 355 (13.1\%) were hot only, and 326 (12.0\%) met both thresholds. Forty cells (1.5\%) were also in the upper poverty quartile. The continuous surface retains the full rank information and identifies where canopy need and high surface temperature are most strongly co-located.

\begin{figure*}[t]
\centering
\includegraphics[width=\textwidth]{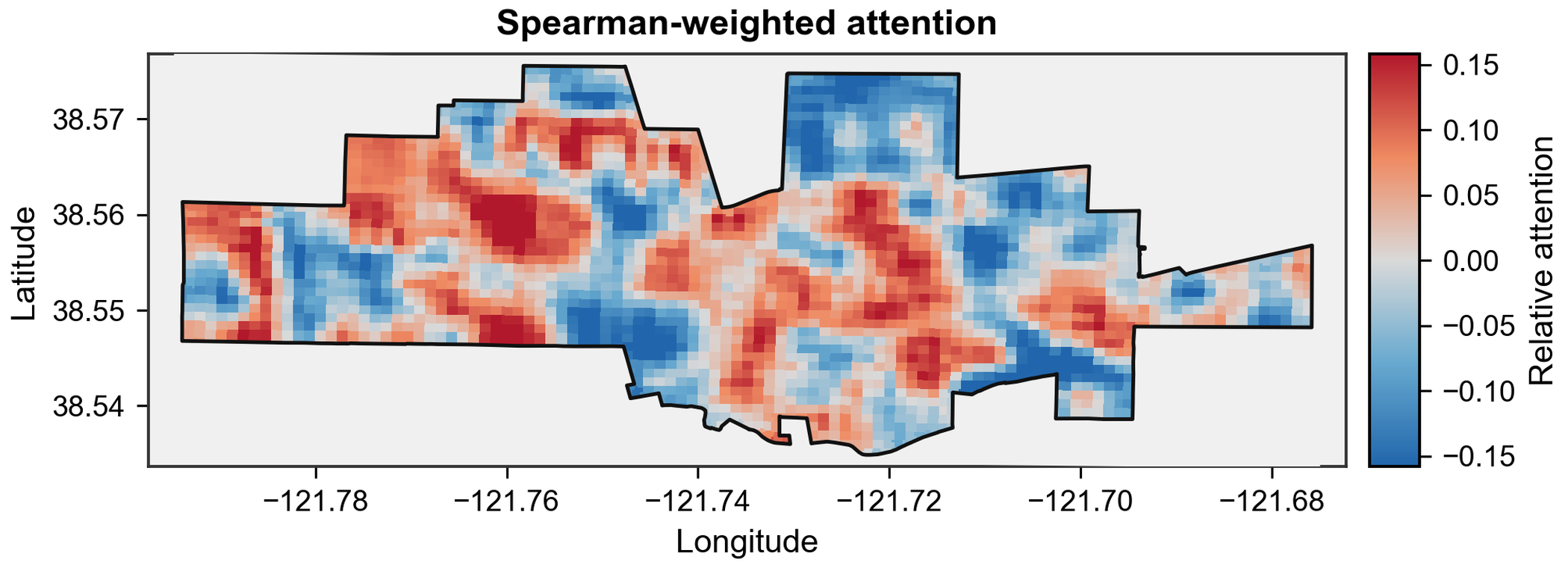}
\caption{Spearman-weighted multi-factor attention surface combining canopy need, residential value per acre, LST, built probability, population density, poverty, income, and distance from parks and bikeways. Red cells receive greater relative attention under the disclosed scenario; the map is not an official equity score.}
\label{fig:weighted}
\end{figure*}

The multi-factor attention surface in Figure~\ref{fig:weighted} incorporates canopy need, residential value per acre, LST, built probability, population density, poverty, household income proxy, and distances to parks and bikeways. Weights ranged from 19.7\% for canopy need and LST to 3.1\% for household income proxy (Table~\ref{tab:weights}). Because every input direction and weight is explicit, the surface can be reviewed or reweighted for alternative planning priorities rather than being treated as a fixed score.

\begin{table*}[t]
\centering
\caption{Disclosed weights and directions for the multi-factor attention surface.}
\label{tab:weights}
\begin{tabular*}{\textwidth}{@{\extracolsep{\fill}}p{0.30\textwidth}rp{0.16\textwidth}p{0.30\textwidth}@{}}
\toprule
\textbf{Layer} & \textbf{Weight} & \textbf{Direction} & \textbf{Role} \\
\midrule
Canopy percentage & 19.7\% & Lower & Primary canopy need; absolute $\rho$ matched to the strongest non-canopy layer. \\
Land-surface temperature & 19.7\% & Higher & Thermal context. \\
Population density & 14.5\% & Higher & Potential exposure context. \\
Dynamic World built probability & 12.3\% & Higher & Built-form context. \\
Residential value per acre & 11.4\% & Lower & Land-market context. \\
Poverty rate & 7.7\% & Higher & Socioeconomic context. \\
Distance to park & 7.3\% & Farther & Amenity-access context. \\
Distance to bikeway & 4.2\% & Farther & Active-travel context. \\
Household income proxy & 3.1\% & Lower & Socioeconomic context. \\
\bottomrule
\end{tabular*}
\begin{minipage}{\textwidth}
\vspace{2pt}
\footnotesize Notes: Weights are normalized absolute Spearman associations with canopy. The surface is a transparent screening scenario, not an official equity score.
\end{minipage}
\end{table*}

\section{Discussion}\label{sec:discussion}

\subsection{A high-confidence optical layer with a defined measurement scope}

The shared-extent comparison clarifies what the workflow measures well. Pixel precision of 0.804, pixel recall of 0.873, and detection-center precision of 0.974 show that the expanded optical product aligns closely with the LiDAR-assisted canopy product on the shared municipal extent. The 7.71\,km$^2$ optical surface is near parity with the 7.11\,km$^2$ structural reference (IoU 0.719). This supports a reproducible optical canopy layer that can be regenerated from routinely available NAIP imagery while retaining an explicit crown-anchored detection chain. Reporting pixel area and candidate-center agreement separately is more informative than compressing the comparison into one accuracy number \citep{weinstein2021benchmark}.

Remaining differences in mapped extent reflect distinct sensing bases. The LiDAR-assisted product incorporates vertical structure, whereas the optical workflow begins with visually identifiable proposals, refines crowns with SAM and NDVI, and then applies a transparent NDVI--texture infill for canopy between and beyond detected crowns. Conversely, the crown-anchored rule limits broad spectral confusion with lawns or other low vegetation. For applications that require total canopy accounting, the structural and municipal products remain primary; for rapid update, candidate review, and cross-layer screening, the optical product provides an auditable complement.

The modular architecture is an additional strength. DeepForest supplies object proposals, NDVI provides a transparent vegetation gate, and SAM refines visible shape. NIR enters through the NDVI calculation and within-box union rather than through SAM itself. The same architecture can be extended through logical-OR window merging, locally annotated validation tiles, and stage-wise ablation, enabling future studies to quantify the contribution of each module while preserving the current workflow as a reproducible baseline.

\subsection{Thermal and street relationships}

The inverse canopy--LST relationship was consistent across bivariate, partial, and spatial-lag analyses. In the preferred model, 1\,\textdegree{}C higher surface temperature was associated with approximately 3.00 percentage points lower canopy after adjustment for built probability, population density, poverty, and neighboring canopy. This agrees directionally with studies showing scale-dependent thermal benefits of tree cover \citep{ziter2019scale,schwaab2021trees} and with recent Geocarto International work emphasizing spatially explicit heat diagnostics \citep{chen2025heat}. Because Landsat represents surface temperature at one overpass, the estimate is interpreted as a robust spatial association rather than a causal cooling effect.

The road analysis adds a different planning dimension. Nearly half of candidate crowns were within 15\,m of a road, but road-area percentage was unrelated to canopy. The finding suggests that the geometry of streets and street-adjacent planting matters more than the amount of road surface in a 100\,m cell. Mapping functional class, pedestrian ways, and bikeways separately allows a municipality to ask where existing shade intersects travel corridors and where field verification may reveal planting opportunities or conflicts with utilities, visibility, pavement, or maintenance responsibilities.

\subsection{Distributional context without a predetermined equity narrative}

Davis did not exhibit a simple bivariate pattern in which higher income or higher land value aligned with more optical canopy. This locally specific result should be interpreted alongside the city's mature central neighborhoods, student-oriented housing, institutional land, planned developments, agricultural edges, and greenbelts. The variables also have different spatial supports, and assessed value per acre is a land-market indicator rather than household wealth. The analysis therefore treats the socioeconomic layers as distributional context rather than as direct measures of benefit or causation.

The spatial model sharpened that interpretation. Population density was not significant after adjustment, poverty retained a small positive coefficient, and LST remained strongly negative. These results do not overturn the wider literature documenting canopy inequity across many U.S. cities \citep{schwarz2015money,locke2021segregation,mcdonald2021disparity}. Instead, they demonstrate the importance of local spatial diagnostics and adjusted models. Neighborhood-scale inventory attributes, public/private ownership, tree condition, and community priorities would add the information needed for program design.

\subsection{Planning relevance and transferability}

The City of Davis reports 26.2\% canopy in its 2020 assessment, close to the 27.4\% fraction in the USDA/CAL FIRE product on the shared municipal extent \citep{davis2025stateforest}. These structural and municipal products remain the appropriate sources for total-canopy baselines. The optical workflow adds a reproducible, high-resolution layer that can be rerun with future NAIP acquisitions, used to review candidate crowns, and connected consistently to heat, infrastructure, and neighborhood indicators. Its emphasis on repeatable, date-specific imagery parallels satellite time-series studies that use seasonal Sentinel-2 trajectories to detect vegetation stress across agricultural landscapes \citep{narimani2025satellite}.

Figures~\ref{fig:shade} and \ref{fig:weighted} provide two levels of screening. The equal-weight shade surface offers a direct view of cells where low canopy and high LST coincide. The weighted surface broadens the screen to include built, socioeconomic, land-market, and amenity indicators. Neither map replaces site assessment. Water availability, utilities, right-of-way width, species suitability, ownership, maintenance capacity, and community knowledge remain essential. The analytical advantage is transparency: each factor, direction, and weight is visible and can be changed.

The approach is transferable as a protocol, not as a universal parameter set. NAIP availability makes it relevant to many U.S. cities, but detector thresholds, crown appearance, phenology, irrigation, and urban morphology will vary. Related field-scale work with analysis-ready geospatial embeddings likewise shows the value of independent spatial splits and uncertainty-aware inference when transferring deep-learning maps beyond their training locations \citep{narimani2026alphaearth}. A new city should therefore repeat the same-extent reference comparison, retain a local manual sample where possible, and refit spatial models rather than transporting Davis coefficients.

\subsection{Limitations}

The principal limitations concern measurement scope and transferability. The optical product does not provide height, species, condition, stem count, or ownership, and the LiDAR-assisted raster is an area reference rather than object-level ground truth. The frozen run does not include a dual-interpreter manual sample or a full stage-wise ablation; canopy expansion thresholds were tuned on the west half of the city and evaluated on the east half against the reference product. Landsat LST represents one surface-temperature acquisition; source vintages and spatial supports differ; grid results remain scale dependent \citep{fotheringham1991maup}; and the cross-sectional models do not establish causation. These limitations are localized to the claims they affect and do not alter the observed shared-extent agreement, spatial associations, or utility of the screening workflow.

\section{Conclusions}\label{sec:conclusions}

Using routinely available 0.6\,m NAIP imagery, the DeepForest--NDVI--SAM workflow produced a citywide crown-anchored canopy layer and 11{,}741 candidate crowns for Davis. On the shared 25.92\,km$^2$ municipal extent, the optical layer mapped 7.71\,km$^2$ (29.8\% of the city) with IoU 0.719 versus the 2022 USDA/CAL FIRE LiDAR-assisted product, and 97.4\% of candidate centers coincided with reference canopy. These metrics define a high-confidence, updateable optical product whose role is complementary to structural and municipal inventories.

The mapped layer supported a coherent GIS analysis. Approximately 49\% of candidates lay within 15\,m of a road. Canopy was inversely associated with Landsat LST in bivariate, partial, and spatial-lag analyses, while socioeconomic patterns were locally contingent after spatial adjustment. A two-factor shade-attention map identified 326 hot, low-canopy cells, and a fully disclosed Spearman-weighted surface provided a broader planning screen. The principal contribution is an auditable progression from optical crown candidates to shared-extent agreement, spatial modeling, and neighborhood screening.

\section*{Acknowledgments}

The authors acknowledge the public agencies and open-data communities that maintain the imagery, environmental, infrastructure, and aggregated statistical products used in this study.

\section*{Funding}

This research received no specific grant from any funding agency in the public, commercial, or not-for-profit sectors.

\section*{Disclosure statement}

The authors report no competing interests to declare.

\section*{Declaration of generative AI use}

During the preparation of this work, the authors used ChatGPT to improve grammatical accuracy, refine sentence structure, and enhance visualizations. All AI-generated revisions were thoroughly reviewed and edited by the authors to ensure relevance and accuracy.

\section*{Data availability statement}

All public source products, collection identifiers, and persistent links are listed in Table~\ref{tab:data}. The NAIP mosaic clipped to the Census TIGER municipal boundary, optical canopy mask, crown detections, 100\,m analysis table, and related derived products are openly available in Zenodo at \url{https://doi.org/10.5281/zenodo.21925526} \citep{narimani2026daviscanopydata}. Replication code is available on GitHub at \url{https://github.com/MohammadrezaNarimaniUCDavis/Davis_Urban_Canopy_GeoAI}.

\section*{Ethics statement}

The study uses remote-sensing, environmental, infrastructure, parcel, and aggregated public statistical data. It involves no human participants and no identifiable individual-level records.

\bibliographystyle{IEEEtranN}
\bibliography{references}

\clearpage
\section*{Supplementary figures}

\noindent
\begin{minipage}{\textwidth}
\centering
\includegraphics[width=\textwidth]{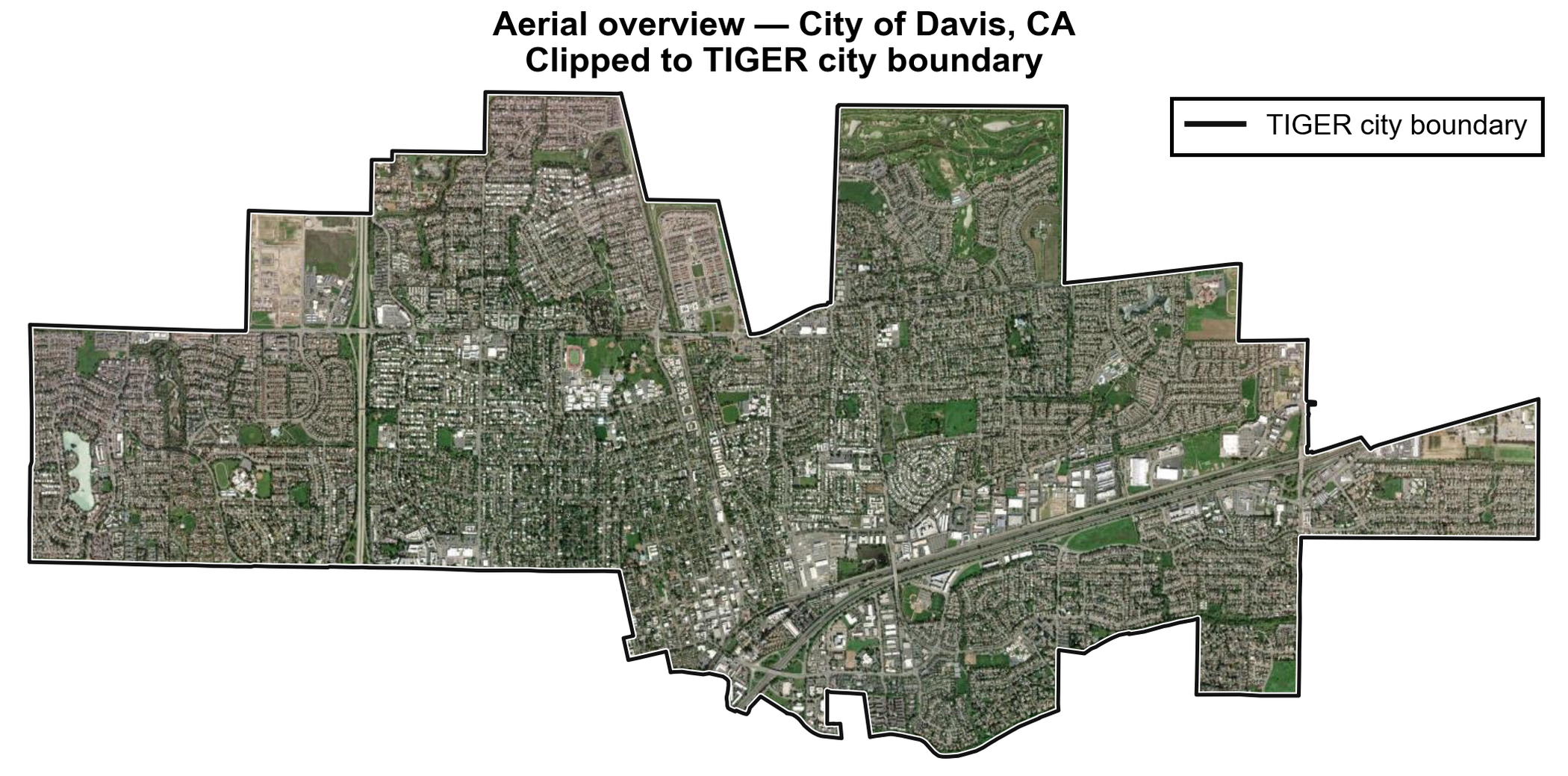}
\captionof{figure}{Citywide 2022 NAIP aerial overview clipped to the Census TIGER municipal boundary.}
\label{fig:s1}
\end{minipage}

\vspace{1.2em}
\noindent
\begin{minipage}{\textwidth}
\centering
\includegraphics[width=\textwidth]{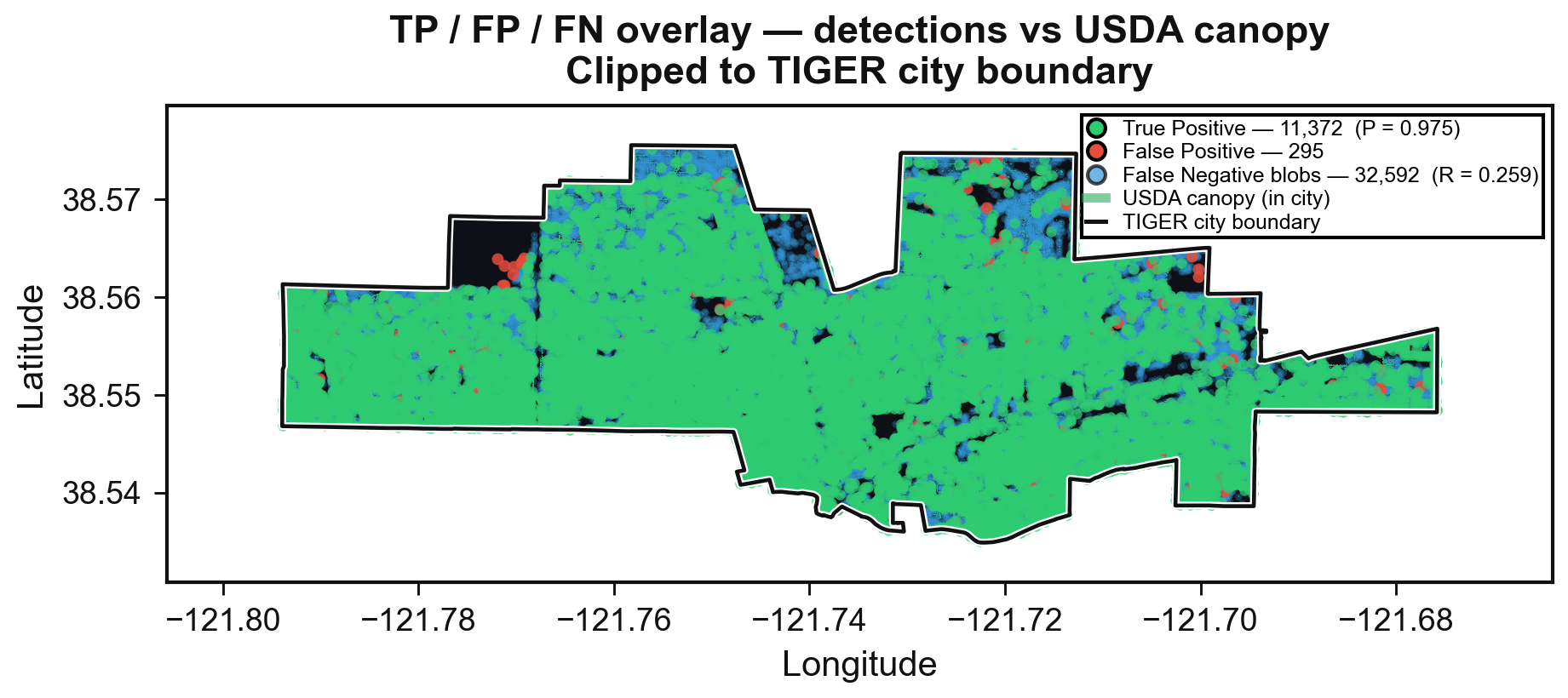}
\captionof{figure}{Detection-center comparison with the USDA/CAL FIRE canopy product on the common city extent. Green points are candidate centers on reference canopy; red points are candidate centers outside the reference canopy. Blue symbols represent reference canopy components not paired with a candidate center under the exploratory overlay procedure. The center-based precision metric is reported in the main text; component counts are not interpreted as object recall.}
\label{fig:s2}
\end{minipage}

\clearpage
\noindent
\begin{minipage}{\textwidth}
\centering
\includegraphics[width=\textwidth]{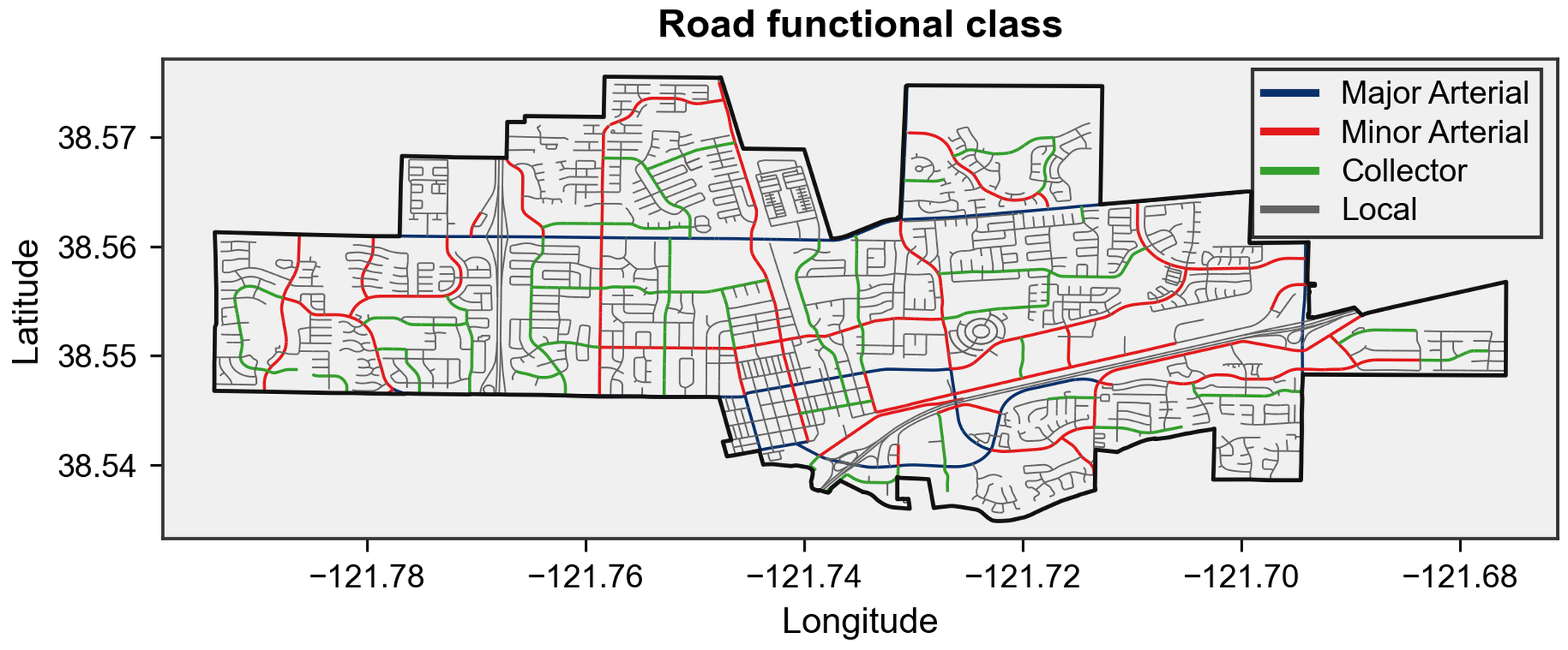}
\captionof{figure}{Yolo GIS road functional classes within Davis: major arterial, minor arterial, collector, and local roads.}
\label{fig:s3}
\end{minipage}

\vspace{1.2em}
\noindent
\begin{minipage}{\textwidth}
\centering
\includegraphics[width=\textwidth]{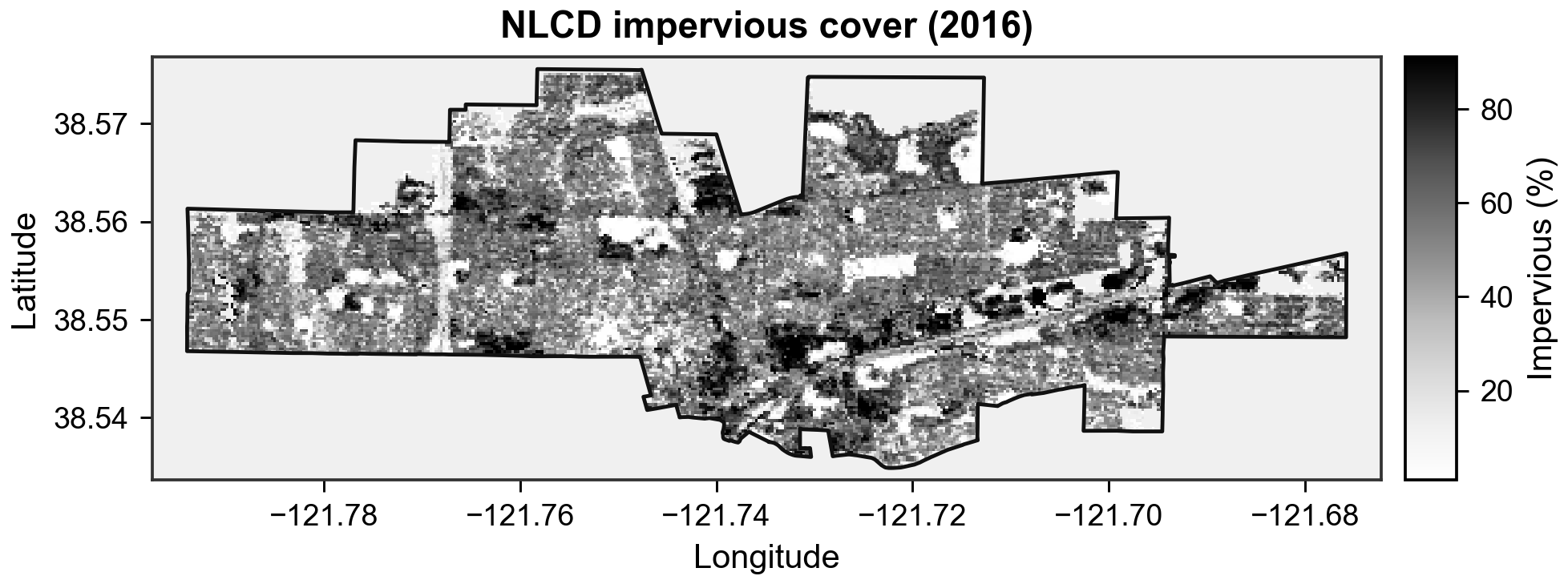}
\captionof{figure}{NLCD 2016 percent impervious cover. This layer predates the 2022 canopy mapping and is presented only as a sensitivity and visual-context product.}
\label{fig:s4}
\end{minipage}

\vspace{1.2em}
\noindent
\begin{minipage}{\textwidth}
\centering
\includegraphics[width=\textwidth]{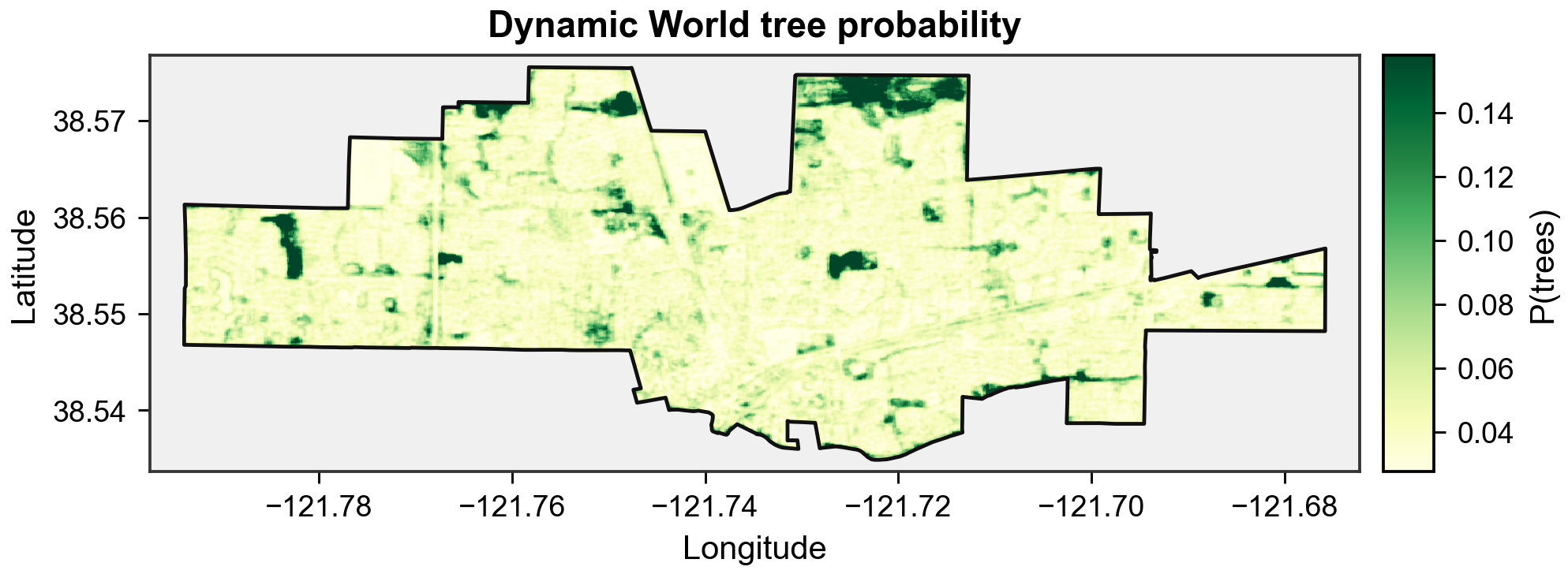}
\captionof{figure}{Dynamic World tree probability for 2022, shown as an independent 10\,m vegetation-context layer rather than an individual-crown reference. Dynamic World data: Google and World Resources Institute, CC BY 4.0.}
\label{fig:s5}
\end{minipage}

\end{document}